\RequirePackage{fix-cm}
\documentclass[pdftex,twocolumn,epjc3]{svjour3}
\smartqed  
\RequirePackage{graphicx}
\RequirePackage[colorlinks,citecolor=blue,urlcolor=blue,linkcolor=blue]{hyperref}
\usepackage{amssymb}     
\usepackage{amsmath}
\usepackage{color}
\usepackage{multirow}
\usepackage{cite}
\usepackage[T1]{fontenc}
\usepackage{upgreek}
\newcommand{\ctsper}      {cts/(keV$\cdot$kg$\cdot$yr)}

\newcommand{\tctsper}     {{$10^{-2}$~cts/(keV$\cdot$kg$\cdot$yr)}}

\newcommand{\pIIbi}       {{$10^{-3}$~cts/(keV$\cdot$kg$\cdot$yr)}}

\newcommand{\kgy}         {{kg$\cdot$yr}}

\newcommand{\mum}         {{$\upmu$m}}
\newcommand{\mus}         {{$\upmu$s}}

\newcommand{\ga}          {$\gamma$}

\newcommand{\qbb}         {{$Q_{\beta\beta}$}}
\newcommand{\thalfzero}   {${T^{0\nu}_{1/2}}$}

\newcommand{\onbb}        {{$0\nu\beta\beta$}}

\newcommand{\etal}        {\textit{et al.}}

\newcommand{\gerda}       {\textsc{Gerda}}
\newcommand{\infn}       {\textsc{Infn}}

\newcommand{\lngs}        {{\mbox{\textsc{Lngs}}}}

\newcommand{\macro}        {{\mbox{\textsc{Macro}}}}
\newcommand{\WT}          {water tank}

\newcommand{\borexino}    {\mbox{\textsc{Borexino}}}

\newcommand{\mage}        {\textsc{MaGe}}

\newcommand{\gesix}       {{$^{76}$Ge}}

\newcommand{\geenr}       {{$^{\rm enr}$Ge}}          
\newcommand{\genat}       {{$^{\rm nat}$Ge}}

\newcommand{\exposure}    {\mbox{$\cal E$}}

\journalname{Eur. Phys. J. C}
\begin{document}

\title{The Performance of the Muon Veto of the \textsc{Gerda}
   Experiment}

\author{
K. Freund\thanksref{TU,nowMaxment} \and
R. Falkenstein\thanksref{TU} \and
P. Grabmayr\thanksref{TU,corrauthor} \and
A. Hegai\thanksref{TU}  \and
J. Jochum\thanksref{TU} \and
M. Knapp\thanksref{TU,nowAreva} \and
B. Lubsandorzhiev\thanksref{TU,INR} \and
F. Ritter\thanksref{TU,nowBosch} \and
C. Schmitt\thanksref{TU} \and
A.-K. Sch{\"u}tz\thanksref{TU}\and
I. Jitnikov\thanksref{JINR}\and
E. Shevchik\thanksref{JINR}\and
M. Shirchenko\thanksref{JINR}\and
D. Zinatulina\thanksref{JINR}
}
\thankstext{nowMaxment}{\emph{present address:} maxment GmbH, Germany}
\thankstext{nowAreva}{\emph{present address:} Areva, France}
\thankstext{nowBosch}{\emph{present address:} Bosch GmbH, Germany}
\thankstext{corrauthor}{\emph{Correspondence}, email: grabmayr@uni-tuebingen.de}
\institute{%
Physikalisches Institut, Eberhard Karls Universit{\"a}t T{\"u}bingen,
    T{\"u}bingen, Germany\label{TU} \and
Institute for Nuclear Research of the Russian Academy of Sciences,
    Moscow, Russia\label{INR} \and
Joint Institute for Nuclear Research, Dubna, Russia\label{JINR}
}\date{Received: date / Accepted: date}
%
%
\maketitle
\begin{abstract}
Low background experiments need a suppression of cosmogenically induced
events. The \gerda\ experiment located at \lngs\ is searching for the
\onbb\ decay of \gesix. It is equipped with an active muon veto the main part
of which is a water Cherenkov veto with 66~PMTs in the \WT\ surrounding the
\gerda\ cryostat.  With this system 806 live days have been recorded, 491 days
were combined muon-germanium data. A muon detection efficiency of
$\varepsilon_{\upmu d}=(99.935\pm0.015)$\,\% was found in a Monte Carlo
simulation for the muons depositing energy in the germanium detectors. By
examining coincident muon-germanium events a rejection efficiency of
$\varepsilon_{\upmu r}=(99.2_{-0.4}^{+0.3})$\,\% was found. Without veto
condition the muons by themselves would cause a background index of
$\textrm{BI}_{\upmu}=(3.16 \pm 0.85)\times10^{-3}$ \ctsper\ at \qbb.
\keywords{ cosmic muons \and water Cherenkov detectors \and scintillation
  detectors \and data analysis}
 \PACS{95.85.Ry \and 29.40.Ka \and 29.40.Mc   \and 29.85.Fj}
\end{abstract}
%
\section{Introduction\label{sec:intro}}
Muons may cause a substantial background to rare event searches like
\gerda\ (\textbf{Ger}manium \textbf{D}etector \textbf{A}rray) by generating
counts in the region of interest (ROI) either through direct energy deposition
in the detectors or through e.g. decay radiation of spallation products. The
\gerda\ experiment is searching for the neutrinoless double beta (\onbb) decay
of $^{76}$Ge~\cite{gerda_tec, gerda_prl}. The experimental signature of the
\onbb\ decay is a peak at \qbb, the $Q$ value of the decay.

\gerda\ was constructed in the underground laboratory of Laboratori Nazionali
del Gran Sasso (\lngs) of \infn\ in Italy, which offers an overburden of 3500
meter water equivalent (m.w.e.) of rock and hence a reduction of the muon flux
by a factor of $\sim$10$^6$ to a rate of
$\sim$3.4$\times10^{-4}$/(s$\cdot$m$^2$). This remaining muon flux however is
sufficient to cause a non-negligible background in the region of interest
around \qbb=2039~keV when increasing the sensitivity beyond
\thalfzero$>$$10^{25}$~yr or when requesting a background index BI$<$\tctsper.
However, also other analyses profit from the reduced backgrounds (see
e.g. Ref.~\cite{majoron,now}).

The origin of muons at \lngs\ for Phase~I of \gerda\ was twofold. Firstly, the
majority of the detectable mu\-ons are produced cosmogenically. Spectrum and
angular distribution of the muons are both altered by the profile of the rock
overburden and have been measured for \lngs\ with high
precision~\cite{macro_mu_93}. These muons have an average energy of $\langle
E_\upmu\rangle=270$~GeV. Secondly, a source for muons was the \textsc{Cngs}
neutrino beam from \textsc{Cern}~\cite{cngs13} which created muons via
e.g. $\nu_\upmu + d \rightarrow \upmu^- + u$ reactions in the vicinity of the
detector. This contribution amounted to 2.2\,\% of the total muon flux in
\gerda. As the \textsc{Cngs} was shut down after 2012 the future Phase~II of
\gerda\ will be unaffected. In order to reduce muon induced background, a muon
veto comprised of a water Cherenkov veto and a scintillator veto was
implemented to tag muons and to use its response as a veto signal in the
\onbb\ analysis.

This paper describes the hardware and setup of the veto and the DAQ
system. The performance of the veto will be presented and compared to Monte
Carlo simulations. Detection and rejection efficiencies for the veto will be
discussed. Parts of this work have been published during the respective PhD
periods~\cite{knapp_phd,ritter_phd,freund_phd,mubuild,mucalib}.
\section{Instrumentation}
Here, a technical description of the apparatus is given. Purpose and function
of both parts of the veto, Cheren\-kov and scintillator panels, is
introduced. The trigger logic, calibration and data acquisition are
summarized.
\subsection{\textsc{Gerda}}
\begin{figure}[t!]
\begin{center}
\includegraphics[width=.45\textwidth]{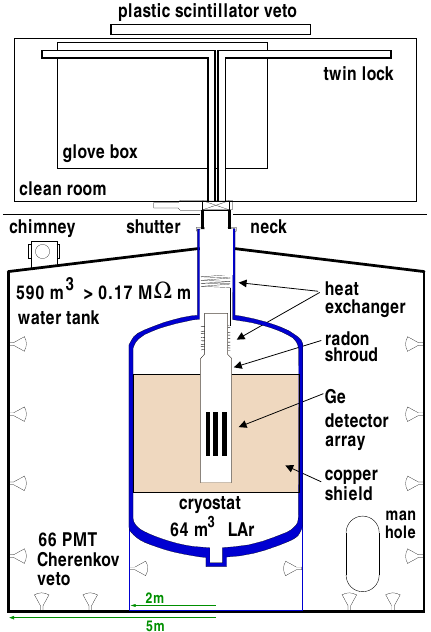}
\caption{  \label{fig:sketch}
         A sketch of the \textsc{Gerda} experiment.}
\end{center}
\end{figure}
In \textsc{Gerda} an array of bare germanium detectors is operated in a
cryostat that contains 64~m$^3$ of liquid argon (LAr) as seen in
Fig.~\ref{fig:sketch}. The cryostat is surrounded by a \WT\ with a diameter of
10~m and a height of 9.4~m that is filled with 590~t ultra-pure water; its
walls are covered with a reflective foil. The cryostat has a connection (neck)
through the \WT\ to the clean room above from which the germanium detectors
are lowered into the LAr. Both \WT\ and cryostat are part of the innovative
shielding design of \gerda\ where low-Z materials are used in order to reduce
cosmogenic activation~\cite{gerda_tec,heusser}.

The \textsc{Gerda} muon veto consists of two independent parts which are read
out by the same DAQ. The first and main part is a water Cherenkov veto that
detects the Cherenkov radiation of the traversing muons. Thus, the \WT\ is
instrumented with 66 encapsulated photomultipliers (PMTs). The second part of
the muon veto is comprised of plastic scintillator panels. Since the
Cherenkov veto offers reduced tagging capability at or around the neck of the
cryostat, plastic panels were placed on top of the clean room in order to
close this weak spot in the Cherenkov veto.
\subsection{The Cherenkov veto}
\begin{figure*}[t!]
\begin{center}
    \includegraphics[width=.48\textwidth]{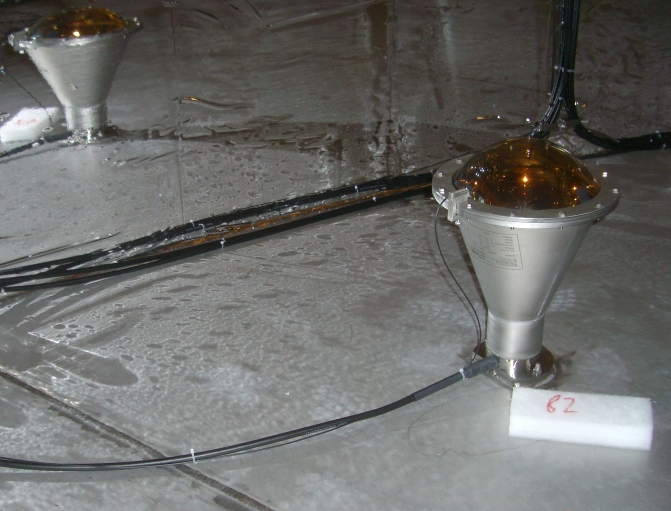}~
    \includegraphics[width=.48\textwidth]{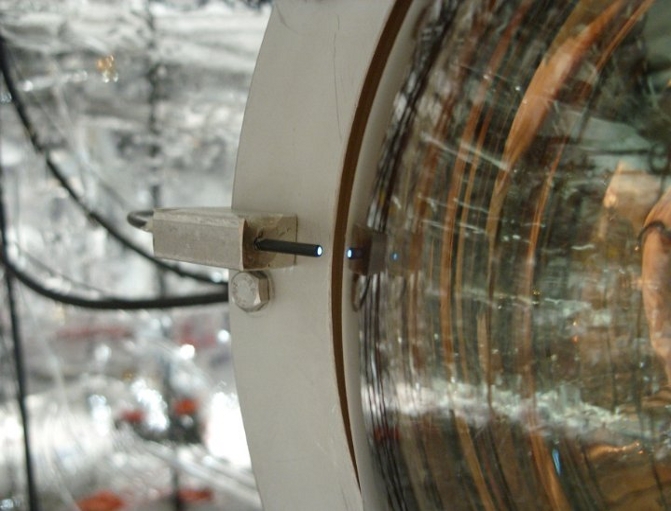}\\
    \includegraphics[width=.48\textwidth]{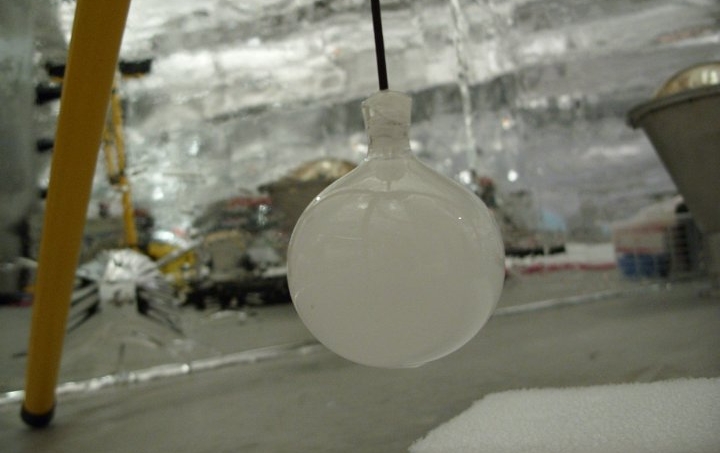}~
    \includegraphics[width=.48\textwidth]{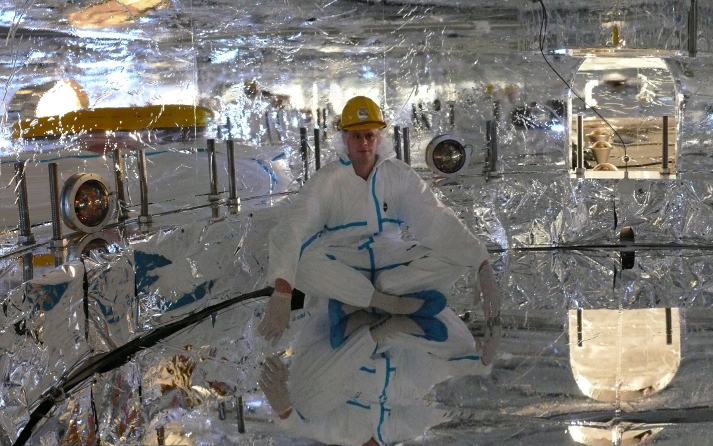}
  \caption{   \label{fig:vetopix}
      Images of the components of the Cherenkov veto during the
       installation. Top, left: a capsule mounted on the floor; top, right: a
       capsule on the wall; bottom, left: a diffuser ball; bottom, right: PMTs
       and Daylighting Film in ``pillbox'' below the cryostat
}
\end{center}
\end{figure*}

Below the water level there are 66 PMTs (8'' size) of type 9354KB/9350KB by
ETL~\cite{935x} installed. In order to protect the PMTs from the surrounding
water, each PMT is housed in a stainless-steel encapsulation which is further
developed from the design of the \borexino\ capsules~\cite{bor_cap}. The
capsule of low radioactivity stainless-steel~\cite{raff} is closed with a
custom made PET cap and is filled with IR spectroscopy oil~\cite{sigma}. The
oil keeps the optical transition between PET cap and PMT as smooth as
possible. Thus, efficiency losses by total internal reflections are
minimized. The bottom of the base of the PMT is encased in polyurethane
\cite{electro} and sealed with silicon gel~\cite{wacker}. The lower part of
the PMT, i.e. the dynode structure, is protected from magnetic fields by a
cone of $\upmu$-metal~\cite{mumet}. Each capsule is equipped with an optical
fibre for optical calibration pulses from outside the \WT. To keep the number
of cables and connectors within the clean water at a minimum an underwater
high-voltage cable~\cite{jowo} connects the PMT to a signal splitter outside
of the \WT. This band-pass separates the signal from the high voltage
(HV). The former is digitized and stored on disk upon a trigger
(sec.~\ref{sec:daq}). The HV is supplied by a HV multichannel system by
CAEN~\cite{1527} equipped with six 12-channel boards \cite{1733}.
\begin{figure}[t]
\begin{center}
  \includegraphics[width=.45\textwidth]{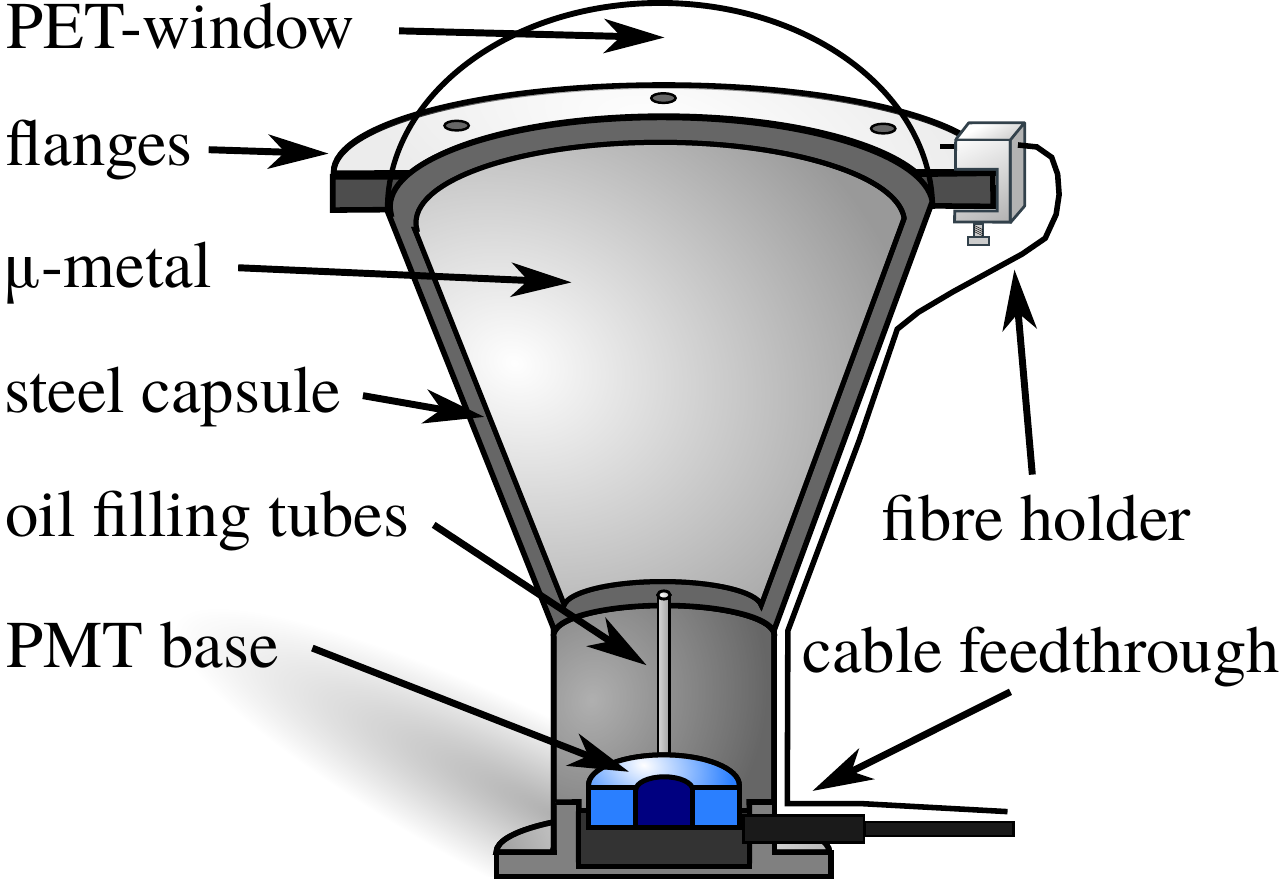}
  \caption{\label{fig:capsule}
      Parts of a PMT capsule
}
\end{center}
\end{figure}
 A sketch of
the capsule is shown in Fig.~\ref{fig:capsule} and images of the components of
the Cherenkov veto are shown in Fig.~\ref{fig:vetopix}.

The water in the \gerda\ \WT\ was provided initiallly by the \textsc{Borexino}
water plant. The purification system which consists of filters, de-ionizers
and an osmosis unit is running with 2.4~m$^3$/hour and keeps the water at the
ultra-pure level of of $>0.17$~M$\Omega$~m~\cite{water}.

\begin{figure}[h!b]
\begin{center}
\includegraphics[width=.46\textwidth]{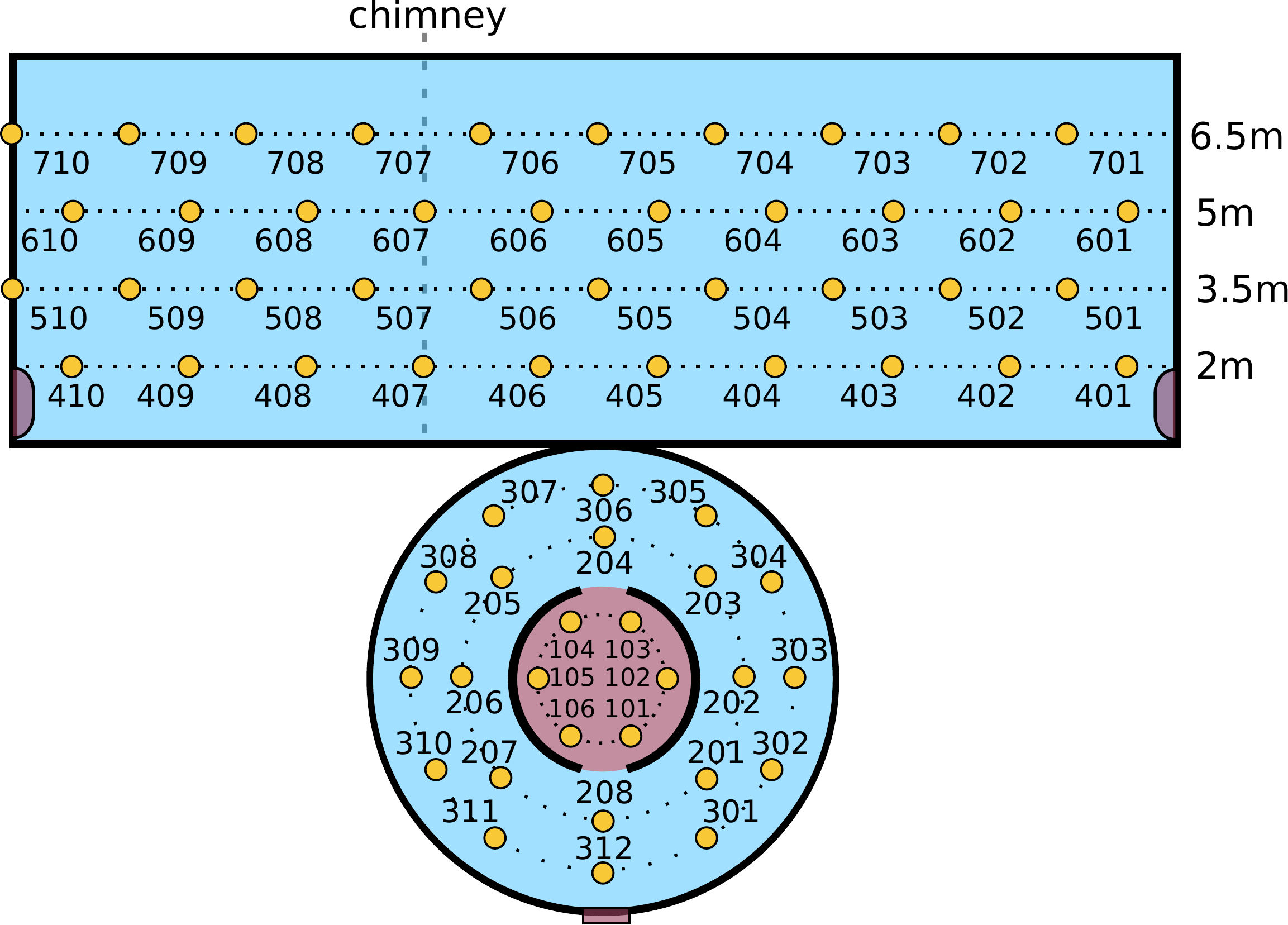}
\caption{  \label{fig:distrib}
        A sketch of the PMT distribution inside the \gerda\ \WT. The violet
        area on the bottom plate signifies the circumference of the cryostat
        and the gaps show the position of manholes into the pillbox. The other
        three violet areas show the location of the manhole into the
        cryostat. All cables and fibres are lead out of the \WT\ through three
        50~cm flanges mounted in a ``chimney'' above the water level (
        Fig.~\ref{fig:sketch})
}
\end{center}
\end{figure}
The PMTs are arranged in seven rings in the \WT, their distribution is shown
in Fig.~\ref{fig:distrib}. One ring of six PMTs (101-106) is pointing inside
the separate volume under the cryostat (commonly referred to as ``pillbox''),
two rings of eight PMTs (201-208) and 12 PMTs (301-312) respectively are
placed on the bottom of the \WT\ looking upwards and four rings of 10 PMTs are
placed on the wall (401-710), pointing horizontally towards the cryostat. The
number of PMTs and their placement was chosen after an extensive Monte Carlo
study~\cite{knapp_phd}. The PMTs closest to the cryostat, i.e those from the
pillbox and the inner ring on the bottom, were selected according to their
performance and radioactivity (type 9354KB is a low activity module). However,
due to the distance to the germanium detectors and the low overall mass the
radioactivity of the muon veto is negligible compared to the one of the
stainless steel cryostat~\cite{knapp_phd}.

\begin{figure}[t]
\begin{center}
\includegraphics[width=.48\textwidth]{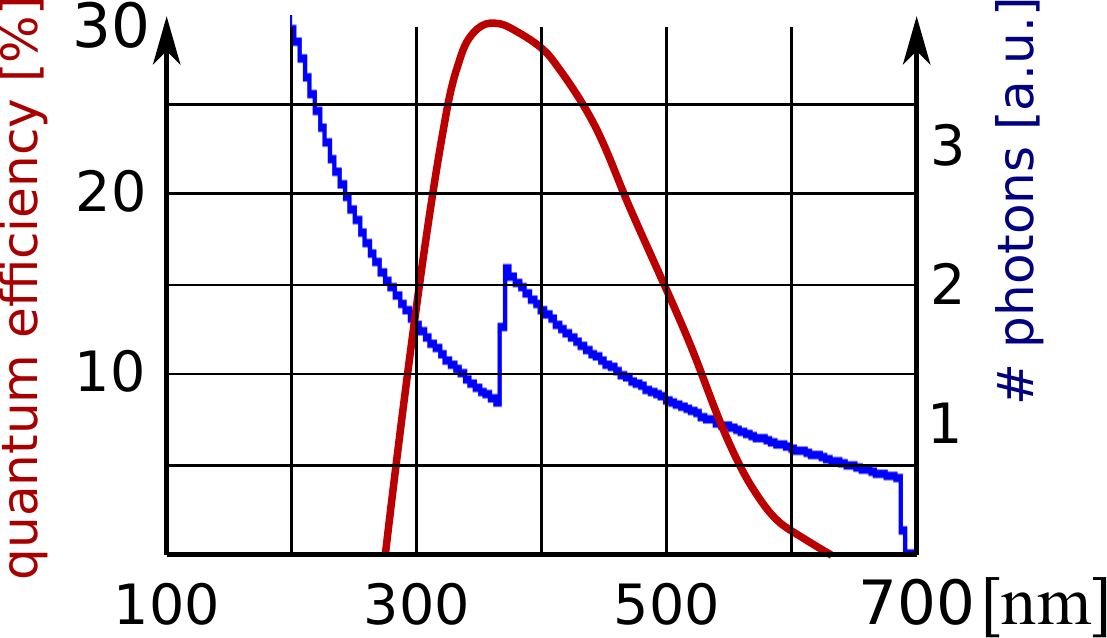}
\caption{   \label{fig:pmteff}
    Efficiency (red curve) of a 9354KB type PMT adapted from
    Ref.~\cite{935x}  and simulated photon spectrum with wave-length
    shifting effect (blue histogram)
}
\end{center}
\end{figure}

The floor and the walls of the \WT\ and of the cryostat are covered with the
reflective Daylighting Film DF2000MA (commonly known as
``VM2000'')~\cite{vm2000} which offers a reflectivity of $>$99\% in order to
increase the light yield of each event.
This comes at the cost of a reduction in tracking capabilities of each
individual muon. The reflective foil does not only increase the collected
light yield of each muon, but it acts as a wave-length shifter as well. It
shifts the predominantly ultra-violet Cherenkov photons to around 400~nm where
the PMTs are most efficient. The efficiency of a 9354KB type PMT and the
wave-length shifting effect implemented in the \textsc{Geant4} simulations are
shown in Fig.~\ref{fig:pmteff}.

\begin{figure}[t]
\begin{center}
  \includegraphics[width=.48\textwidth]{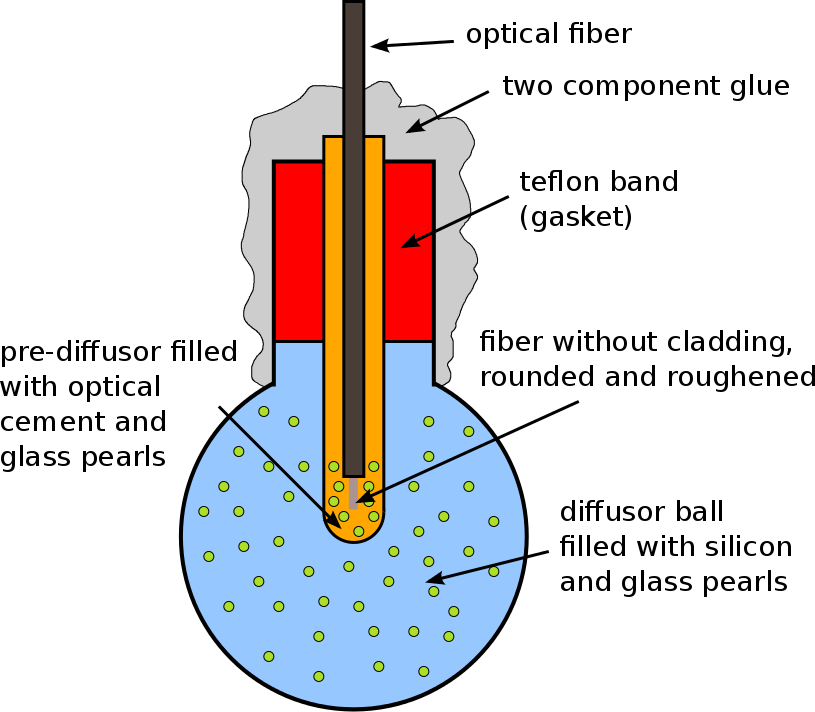}
  \caption{   \label{fig:diffball}
   Schematic drawing of a diffuser ball after
   Ref.~\cite{mucalib}
}
\end{center}
\end{figure}
The PMTs are calibrated regularly with a set of five custom made diffuser
balls shown in Fig.~\ref{fig:diffball} which are constructed to provide a
light source as isotropic as possible. These are glass balls with a diameter
of 4.5~cm that are filled with a mixture of silicon gel~\cite{wacker} and
glass pearls~\cite{pearls} with a diameter of $\sim$50~\mum. An optical fibre
is glued into a small vial inside the ball with a higher content of glass
pearls. The cladding at the end of the fibre is removed and the fibre is
roughened in order to obtain a diffuse light emission. Ultra-fast LEDs outside
of the water tank can be pulsed to illuminate the five diffuser balls
\cite{mucalib}. Four of these balls are distributed in the main \WT\ and a
fifth is placed inside the ``pillbox''. With an appropriate setting of the
LEDs it is possible to illuminate all PMTs simultaneously with single photons
and thus record single-photon responses of every PMT at the same time. As the
diffuser balls are working very well, the calibration fibres attached to every
PMT are currently not in use.
\subsection{Scintillator Veto}
Muons passing through the neck of the cryostat may either traverse a too short
distance in the ``pillbox'' or in the \WT\ to be detected. In order to keep
the muon rejection efficiency as high as possible, a veto of plastic
scintillator panels was conceived and installed.

Each scintillator panel contains a $200\times50\times3$~cm$^3$ sheet of
plastic scintillator based on polystyrol with an addition of PTP (2\,\%) and
POPOP (0.03\,\%)~\cite{scint}. In addition the panels contain optical
fibres~\cite{fibre} on the narrow sides as light-guides, an electronics board
with a trigger and shaper and a PMT. Half of the panels are equipped with PMTs
by Hamamatsu Photonics~\cite{ppmt} and the other half with PMT-085 by
Kvadrotech. The PMT-085 are powered by the same HV supply as the muon veto
PMTs however connecting 3 PMTs to one HV output. The front-end electronics,
the Hamamatsu PMTs and all other components are powered by a custom made power
source with +12~V and $\pm$6~V.

\begin{figure}[t]
\begin{center}
\includegraphics[width=.48\textwidth]{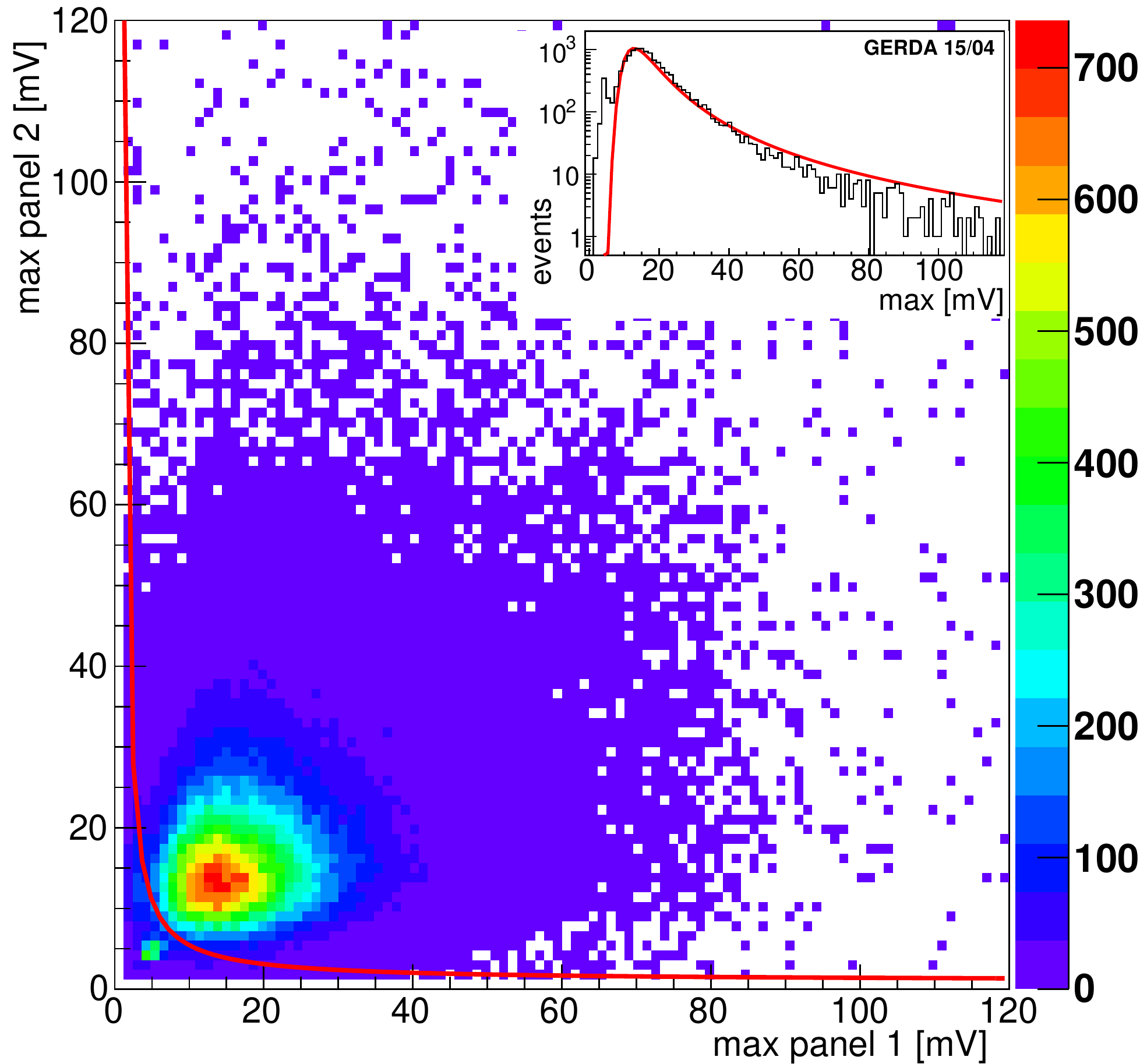}
\caption{   \label{fig:panscat}
     Scatter plot of the pulse height maxima of two panels in a triple stack
     with a  product cut (red line) as given in Eq.~\ref{eq:eq1}. The inset
     shows the pulse height distribution of one panel and a Landau-fit (red)
}
\end{center}
\end{figure}

The panels are arranged in three layers covering an area of $4\times3$~m$^2$
centered over the neck of the cryostat. It was aimed to keep both trigger rate
and thus data volume of the veto as low as possible. Therefore, the trigger is
a triple coincidence of all three layers since this option offers a high
discrimination against non-muonic background events in the panels such as
\ga\ rays from environmental radioactivity. Thus, the scintillator veto
records an almost pure muon sample. Individually, each panel shows a
pulse-height distribution which takes the form of a Landau peak. Despite the
triple coincidence, small \ga\ contamination at low pulse heights
remains. These events can be discarded with a product cut of the form
\begin{equation}(x_1-\rho_1)(x_2-\rho_2)<c\label{eq:eq1}
\end{equation}
for the pulse heights $x_{1,2}$ of any pair of panels in the stack and constants
$\rho_{1,2}$ and $c$. The 2-D pulse height distribution of two panels in a
stack, the applied cut and a spectrum of a single panel with a fit to the
Landau peak is shown in Fig.~\ref{fig:panscat}.

\subsection{Data acquisition\label{sec:daq}}
\begin{figure}[t!]
\begin{center}
  \includegraphics[width=.46\textwidth]{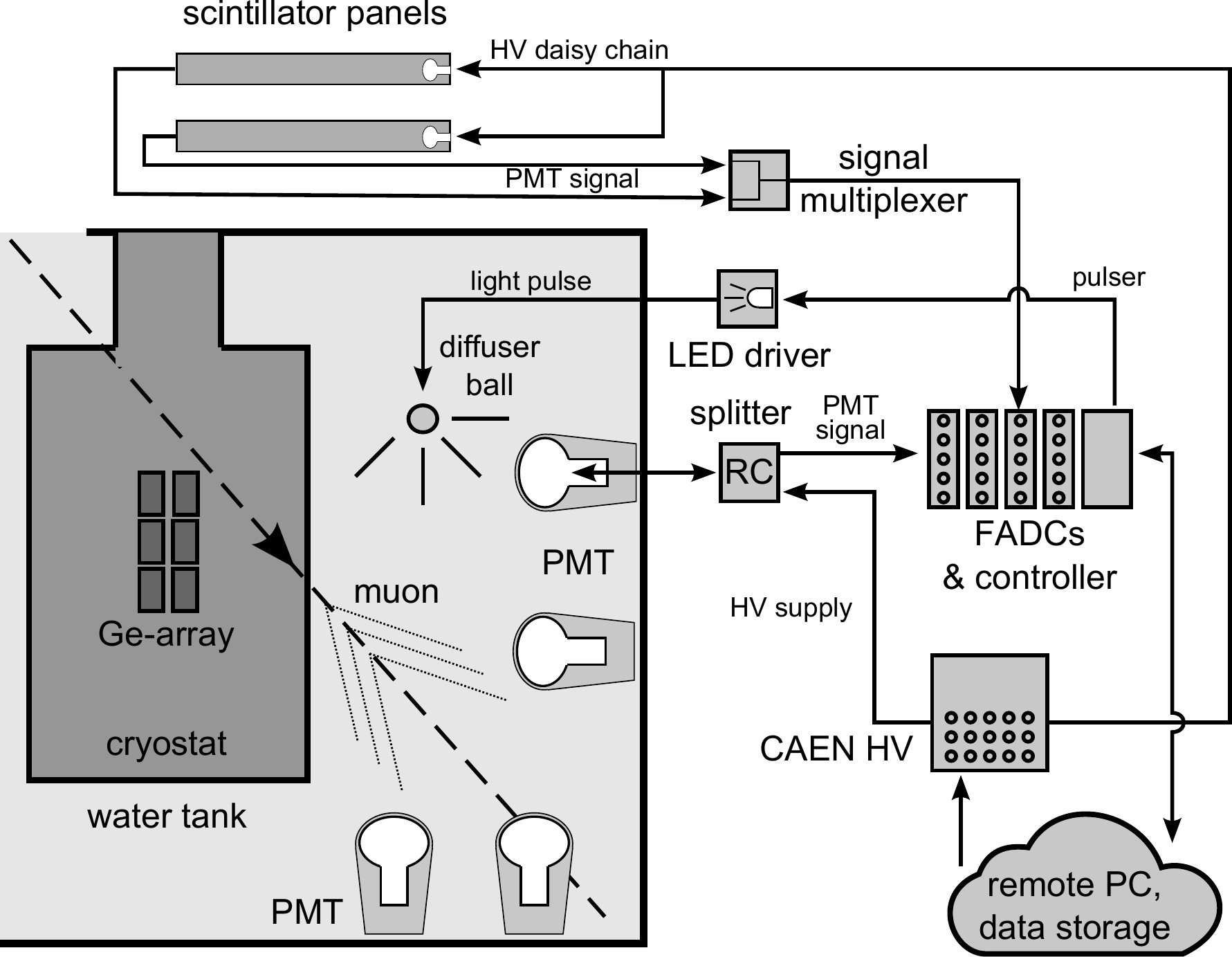}
  \caption{    \label{fig:daqscheme}
         Schematic drawing of the data acquisition (DAQ) setup
}
\end{center}
\end{figure}

The entire apparatus is read out and operated by a VME data acquisition system
(DAQ) which is almost identical to the one of the germanium DAQ (see
Ref.~\cite{gerda_tec} for details). Ten Flash-ADCs~\cite{struck} with 8
channels each digitize the input signals of the Cherenkov veto with 100~MHz.
The signals in each channel are processed by a trapezoidal filter and if the
height exceeds the threshold set to 0.5 photo-electrons (p.e.)  an internal
trigger is generated. Each FADC module has one trigger output which is the
logic OR of the internal triggers of its eight channels.  Thus, the PMT
signals are connected to the input channels in such a way over the FADCs, that
neighboring PMTs are always read out by different FADCs.  This allows the
proper detection of clustered events in which PMTs next to each other have
fired. The final trigger condition is set such that five FADCs must trigger on
at least 0.5 photo-electrons (p.e.) each within 60~ns.  That signal is
realized by a FPGA and it starts the readout of all traces covering a period
of 4~\mus. These are stored together with the time stamp, lately from a GPS
clock.  A schematic drawing of the entire muon veto and its data flow is shown
in Fig.~\ref{fig:daqscheme}.  The trigger signal is furthermore sent to the
germanium DAQ recording it as a redundant but immediate veto information for
the germanium data stream.

The scintillator panels are arranged in three layers of 12 panels each. Three
additional FADCs of the same type digitize the signal of the scintillator
veto. The signals of two non-neighboring panels within a layer are multiplexed
onto one FADC channel using custom made reflection-free modules with an
amplification of -6~dB. Thus the 36 panels occupy only 18 FADC channels such
that each layer is read out by one FADC module.  The panel stack which was hit
can be determined by the unique combination of fired channels. The same
trigger logic as for the Cherenkov veto is applied for the panels, albeit the
trigger window is larger in order to accommodate the much longer output
signals of the panel PMTs because they are shaped with a larger time
constant. For a panel event, all three FADCs (i.e. all three panels in a
stack) need to have fired. Both types of trigger signals are accepted during
data taking.

For a calibration run of the PMTs, the standard data taking is stopped. The
ultra-fast calibration LEDs are activated with a pulser and the LED luminosity
is controlled by a current source in form of a digital-to-analog
converter~\cite{dac}. A separate FADC reads out the pulser signal and triggers
the entire veto for each pulse.

\begin{figure}[t!]
\begin{center}
  \includegraphics[width=.48\textwidth]{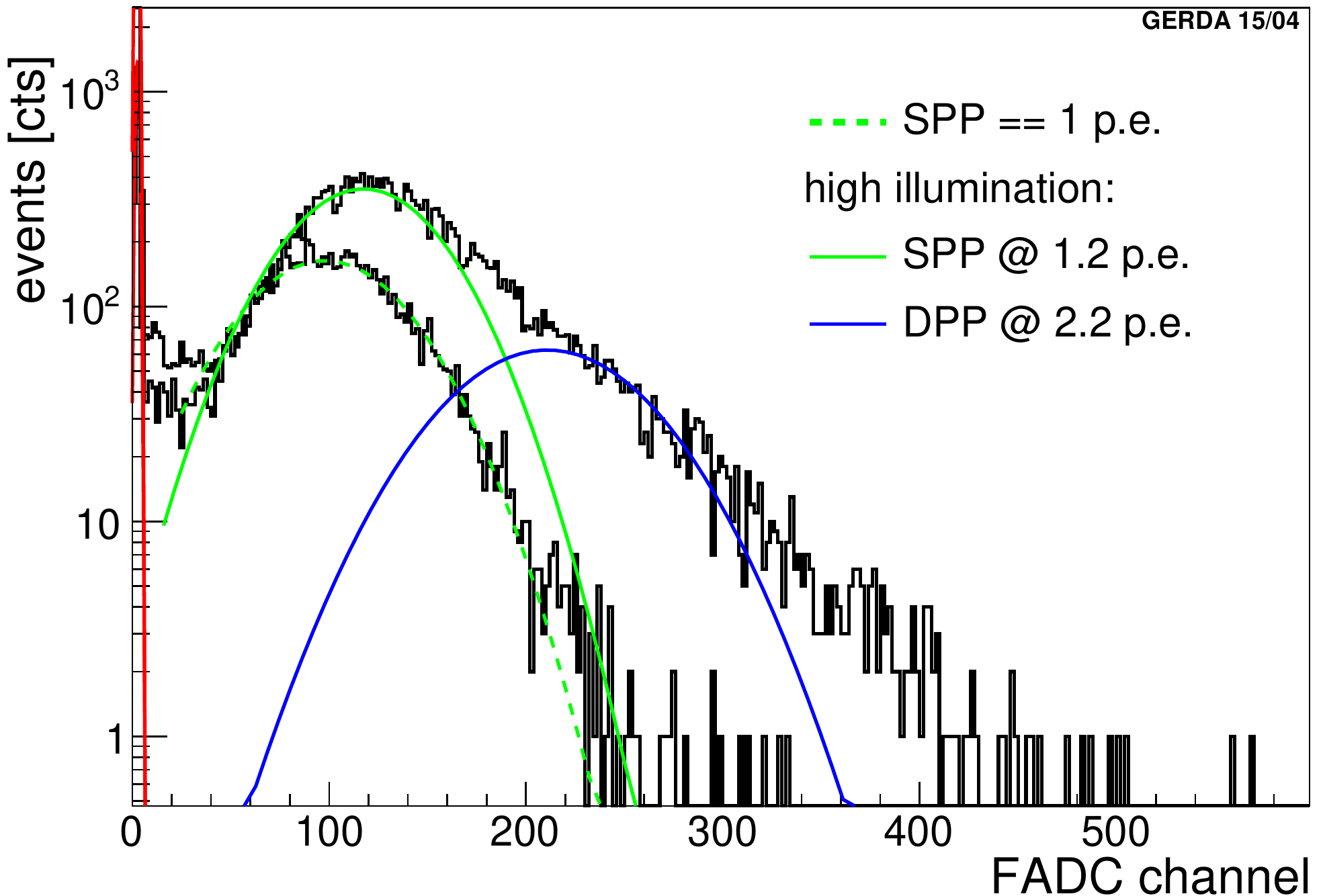}
  \caption{      \label{fig:calib}
       Cherenkov PMT response with different forward voltages of the
       calibration LEDs, i.e. luminosities. With a low luminosity, only the
       single photon peak (SPP, broken green line) is visible, if the
       luminosity is too high, the double photon peak (DPP, blue) emerges as
       well. The pedestal is shown in red
}
\end{center}
\end{figure}

Two calibration spectra can be seen in Fig.~\ref{fig:calib}. One is showing
the conventionally recorded single-photon peak (SPP) set to channel 100, the
other comes from a too bright LED setting. This causes a contamination of the
SPP by the double photon peak (DPP) and a shift of the amplitudes of both
peaks to higher values. The SPP emerges very well in most PMTs and
peak-to-valley ratios between 1.2 and 3.0 are observed.

\section{Veto performance}
\begin{figure}[b]
\begin{center}
  \includegraphics[width=.45\textwidth]{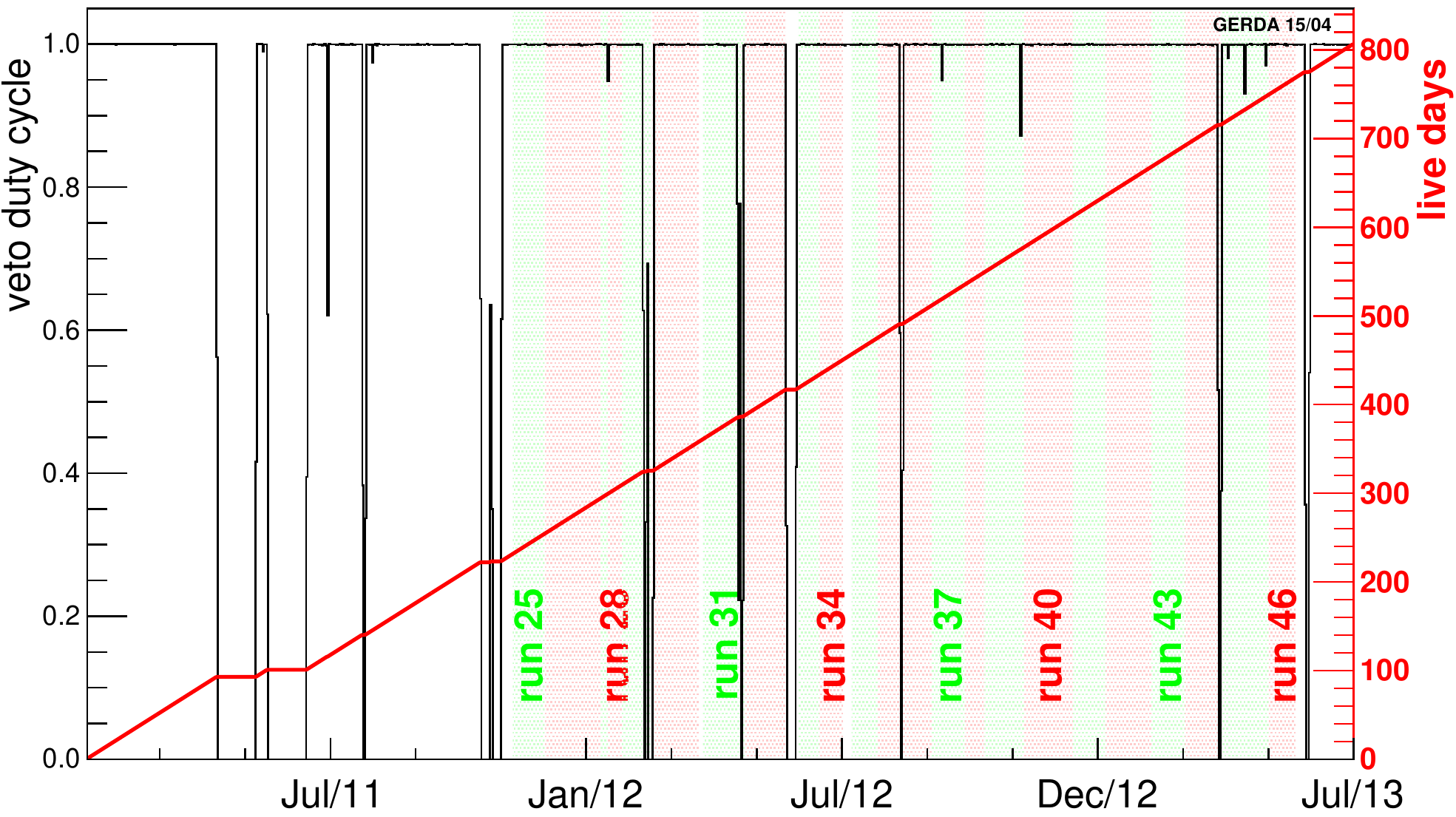}
  \caption{    \label{fig:duty}
       Duty cycle of the muon veto. The veto uptime (black) and accumulated
       live days (red) are marked as well as the \gerda\ physics runs for
       Phase~I (filled light red and green)
}
\end{center}
\end{figure}
\begin{figure}[t!]
\begin{center}
  \includegraphics[width=.49\textwidth]{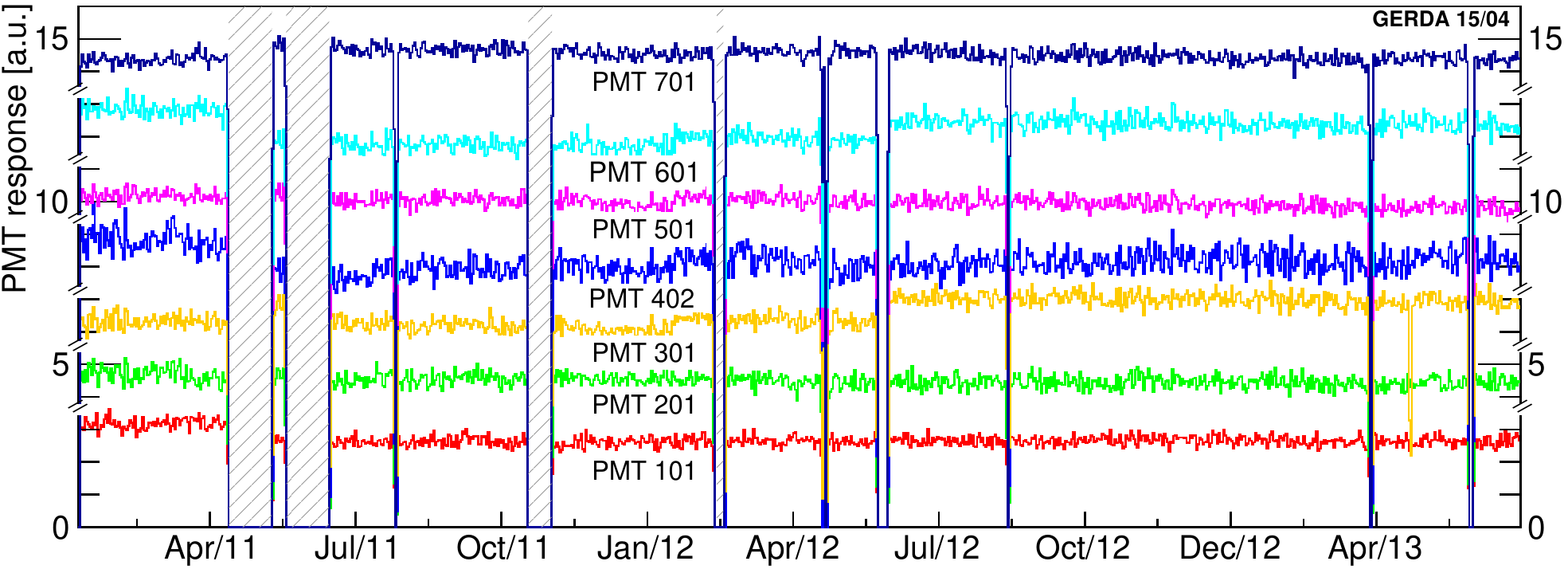}
  \caption{      \label{fig:stabil}
       Summary of the stability of the light output of selected PMTs; note the
       arbitrary offsets. The hatched areas indicate maintenance periods
}
\end{center}
\end{figure}

%
The \textsc{Gerda} muon veto was installed in 2009 and its operation started
in November 2010. During the germanium commissioning runs the panel veto was
installed so that the complete veto was operational at the start of the
physics runs of \textsc{Gerda} in November 2011. Until July 2013, 805.6 live
days have been recorded and 491 days of combined muon-germanium data. The duty
cycle is shown in Fig.~\ref{fig:duty} together with the accumulated live time
(red line). During Phase~I the muon DAQ was only stopped during breaks of the
germanium data taking in order to perform short maintenance work and to
calibrate the PMTs by adjusting the HV of each PMT so that each module shows
the same response to single photons. The PMTs were very stable and since the
beginning only few PMTs needed to be readjusted. As example the daily light
output per day is shown for selected seven PMTs in Fig.~\ref{fig:stabil}. The
offsets are for display only. For e.g. PMT101 there was no readjustment
necessary since May 2011, while PMT301 needed several tunings of the
HV. However, between the breaks the light output remains stable.
During the germanium data taking the muon veto was always fully operational.

During Phase~I two PMTs were lost due to implosion of the tube (PMTs 401 \&
604). The implosion was mostly contained by the encapsulation and no other
PMTs in the vicinity were harmed. The implosions happened in February and
April 2012 and the PMTs were over 10~m apart. Hence a direct influence can be
ruled out. A third PMT was lost right after installation due to a punctured
cable (PMT 305) and a fourth PMT ceased working during the installation and
could be immediately exchanged (PMT 203). In July 2013, the \gerda\ \WT\ was
drained, the veto inspected and two of these PMTs were successfully exchanged
(PMTs 305 \& 401). All other PMTs still work as intended and showed little to
no signs of deterioration.
\subsection{Simulation studies}
The performance of the muon veto was simulated with the
\textsc{Geant4}-based~\cite{geant4} framework for \textsc{Gerda} and
\textsc{Majorana} (\mage)~\cite{MaGe}. First, the simulations were used to find
initial placements and efficiencies~\cite{knapp_phd} and repeated once the
exact geometries of the apparatus were finalized~\cite{freund_phd}. The muon
spectra provided by the \textsc{Macro} experiment were used as input for the
simulation of cosmogenic muons~\cite{macro_mu_93}. The simulations were mainly
used to determine the efficiency of the apparatus in case the muon caused any
energy deposition in the germanium crystals.

For the efficiency of the Cherenkov veto the simulation was undertaken in two
parts. Firstly, muons were simulated with the Cherenkov effect switched
off. The primary vertices of those muons that caused energy deposition in the
germanium detectors were extracted from these events. Secondly, these selected
muons were used in a second simulation with Cherenkov effect enabled. This
two-step procedure was applied in order to accelerate the simulation as only a
minute fraction of the muons interact with the germanium detectors and because
the simulation of optical photons is a resource-demanding procedure. A
detection efficiency for the veto was derived by determining the fraction of
energy depositing muons which caused a trigger signal in the muon veto. The
trigger condition were the same as for the muon veto DAQ described in
Sec.~\ref{sec:daq}. For the entire veto a detection efficiency of
\begin{equation}
\varepsilon_{\upmu d}^\mathrm{MC}= (99.935\pm0.015)\% \label{eq:simeff}
\end{equation}
for muons with energy deposition in the germanium detectors was found in the
simulated data. 

By removing certain PMTs from the efficiency calculations, a veto degradation
(e.g. possibly broken PMTs or malfunctioning FADCs) was
simulated. Even with the first two FADCs removed (14 PMTs in total, two in each
of the seven rings shown in Fig.~\ref{fig:distrib}), the
efficiency is still (99.525$^{+0.025}_{-0.035}$)\%. However when only four PMTs
in the pillbox  are removed the value drops to (97.855$\pm0.065$)\%. The
pillbox PMTs can hence be considered the most critical ones in case of a
break-down. 

The light produced inside the pillbox can illuminate the main \WT\ through two
small manholes. However, the light coming from these two holes is not
sufficient in order to generate a trigger. The insensitivity against a loss of
a few PMTs gives high reliability of the efficiency of the veto against small
variations of trigger conditions or changes in the amplitude of the PMTs.

In an earlier work the efficiency was estimated as $\varepsilon_{\upmu
  d}^\mathrm{MC}=(99.56\pm0.42)$\,\% and thus some what lower despite lower
trigger conditions~\cite{knapp_phd}. The previous simulation neglected the
optical connection between the ``pillbox'' volume and the main
\WT\ completely. Several other simulation studies have been performed like the
veto response to regular cosmogenic muons which are shown and compared to
experimental data in the next sections.
\subsection{Multiplicity and photon spectra}
\begin{figure}[t!]
\begin{center}
\includegraphics[width=.48\textwidth]{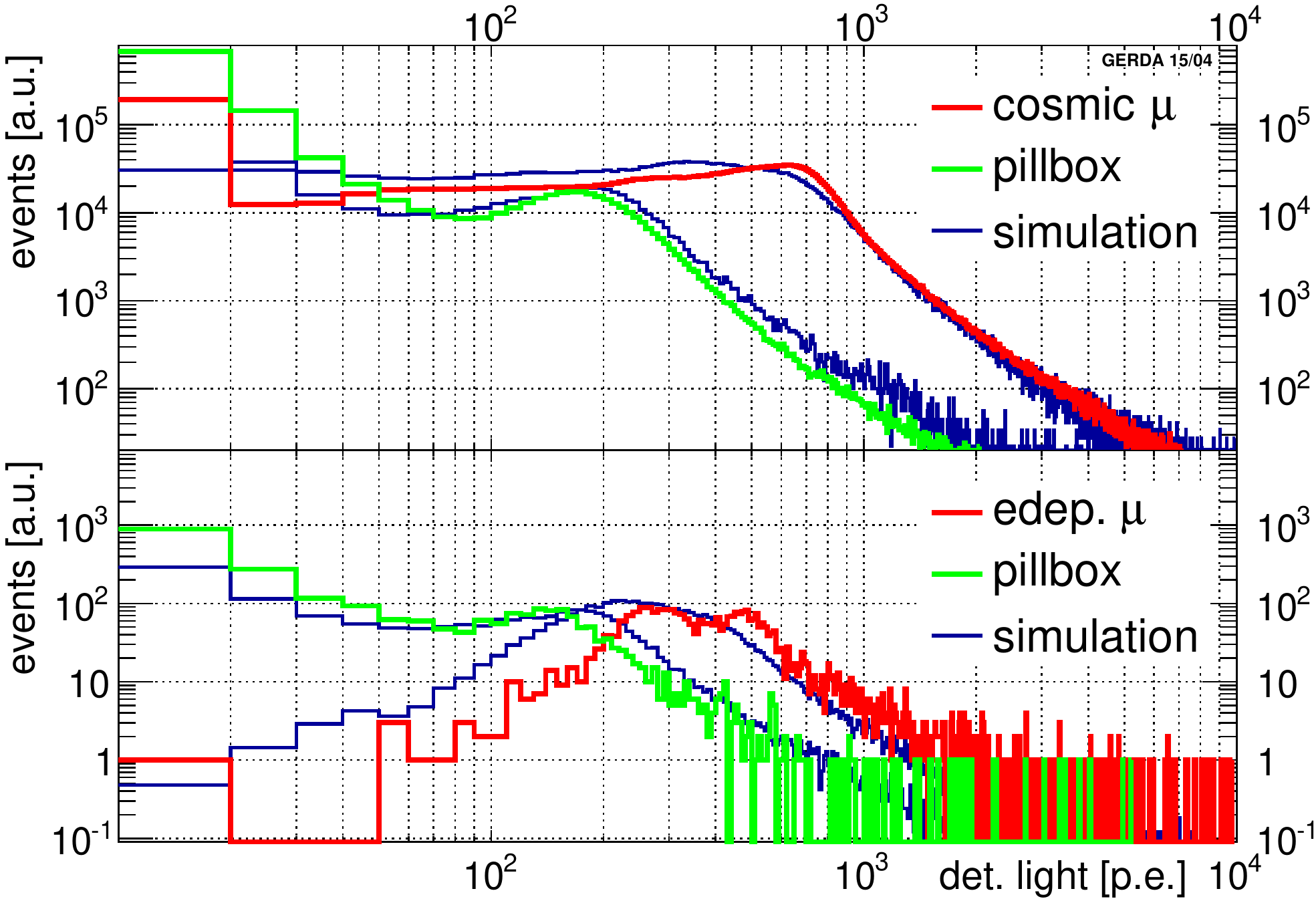}
\caption{     \label{fig:lenght}
       Photo-electron spectra for all cosmogenic muons (top) and those with
       energy deposition in the germanium detectors (bottom). In each panel
       the total recorded p.e. spectrum (red) is compared to the spectrum
       which is recorded just in the pillbox (green). Spectra derived from
       simulations (blue) are normalized to the same exposure
}
\end{center}
\end{figure}%
The light yield of a single muon is determined by observing the total number
of recorded p.e. in all PMTs. For comparison, the pillbox is treated as an
individual volume. In Fig.~\ref{fig:lenght} histograms of the recorded and
simulated events are shown for the period of Phase~I. In the spectra for the
cosmogenic muons, apart from the very low light yield there are maxima at
about 167~p.e. in the pillbox data and about 605~p.e. for the total
spectrum. These broad peaks correspond to the mean traversed distance of 1.8~m
for the pillbox and of 9~m for the \WT\ for muons with a mean incident angle
of $60^{\circ}$. Light from muons in the \WT\ is subject to attenuation, the
attenuation length of photons in water being $\sim$10~m. Assuming a mean
distance of 5~m of the PMTs from the muon track in the \WT\ and 2~m in the
pillbox, muons of any track deposit approximately the same amount of
light. Thus, each muon generates $(115\pm39)$ p.e./m. This is reproduced by
the simulation.

A peak structure is visible in the data for muons which deposit energy in the
germanium detectors as well. The peak for the pillbox is at slightly lower
p.e. values because the average incident zenith angle is lower and thus the
track length shorter. The p.e. spectrum for all PMTs shows a double peak
feature. This is due to the fact that the muon has to cross the cryostat in
order to deposit energy in the germanium. The muon can deposit light in the
\WT\ before and after interacting with the germanium detectors. The higher
peak corresponds to muons which pass the tank twice and the smaller
corresponds to muons that pass the \WT\ just once (e.g. shallow angles close
to the neck of the cryostat). Again, the simulations agree with this even
though the double peak structure is less pronounced.

\begin{figure}[t!]
\begin{center}
\includegraphics[width=.48\textwidth]{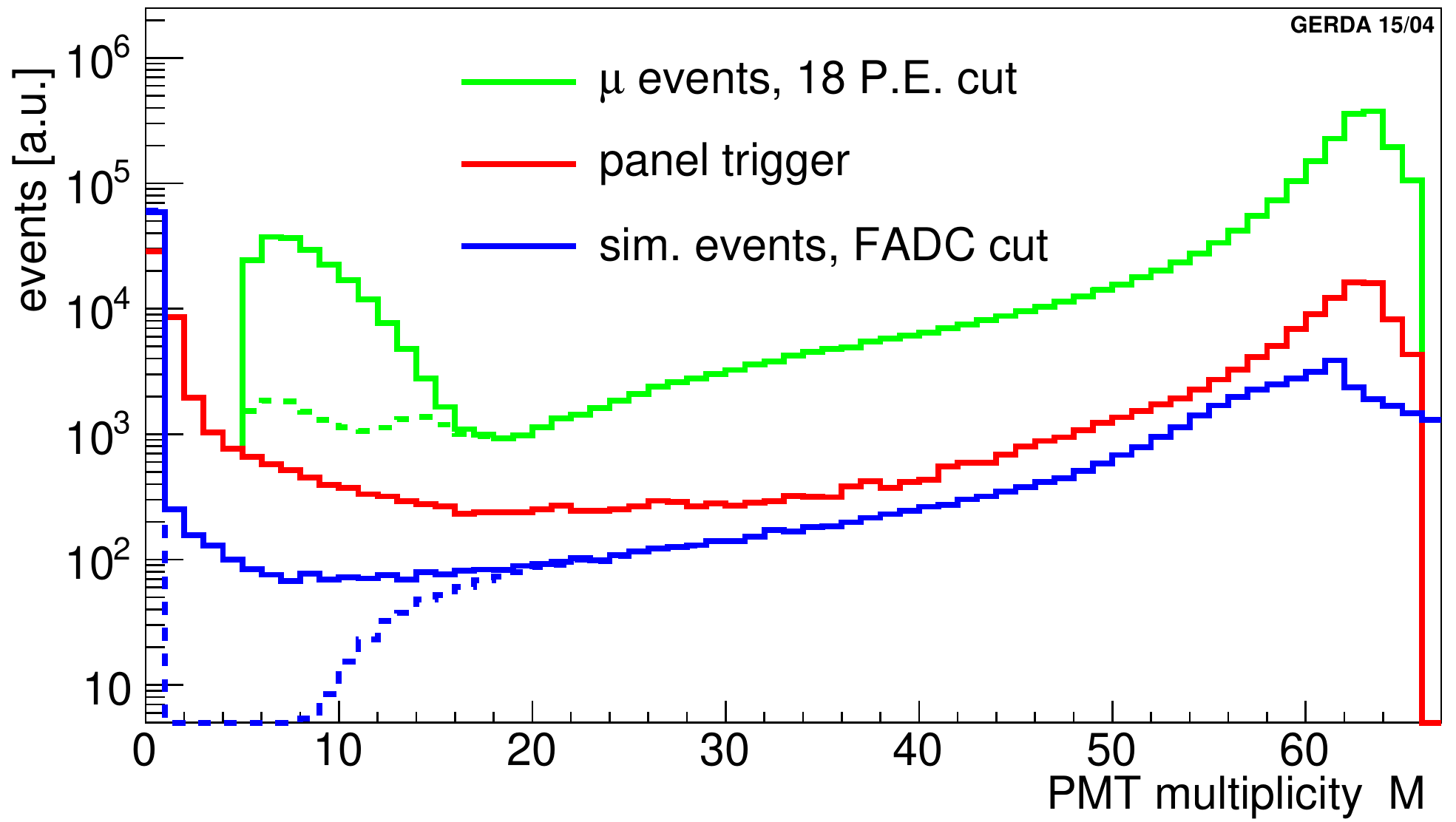}
\caption{       \label{fig:mult}
       PMT multiplicity spectrum of the Cherenkov veto. The multiplicity of
       all events (green) is compared with events where the panels have fired
       as well (red) and with simulated data (blue). 
       Histograms after cuts are marked by dashed lines
}
\end{center}
\end{figure}

Another characteristic of a muon event is the number of fired PMTs. This
multiplicity $M$ is shown in Fig.~\ref{fig:mult} for several classes of muon
events. The spectrum of all measured events (green line) shows a peak at
$M$$>$60 and another one at $M$$<$10. The peak at high $M$ is the regular
response of the veto to muons. This is verified by either the simulated data
(blue line) and a subset of all events in which the panel veto has to have
fired as well (red line). The shape of the spectra is slightly different which
is due to different incident angles (panel trigger) or the lack of random
coincidences at low multiplicities in the simulated data. In addition, four
PMTs were lost in the Cherenkov veto which enhances the peak at $M$$>$60. This
is due to events that would trigger all PMTs are now recorded as having an
$M$=($66-x$), where $x$ is the amount of lost PMTs. Thus, the counts at or
just below 66 are shifted to lower multiplicities.

Only the measured spectrum from the Cherenkov trigger shows the low
multiplicity enhancement, which is characterized by not only a low number of
fired PMTs but also an unusually low amount of recorded p.e. With a cut of
18~p.e.  this enhancement can be almost entirely suppressed. This is the
standard cut condition for the Cherenkov veto data. The source of this
enhancement is discussed in Sec.~\ref{sec:bump}. A cut which emulates the
trigger condition implemented in the DAQ described in Sec.~\ref{sec:daq}
(``FADC cut'') was applied to the simulated data. The resulting spectrum
(dashed blue line) indicates the behavior of the Cherenkov veto without
unphysical events at low $M$ like random coincidences or the low multiplicity
enhancement.
\subsection{Coincident muon-germanium events}
The muon veto and the germanium systems have been operational in common during
Phase~I of \textsc{Gerda} over a time of 491~d. During this period an exposure
of $\exposure=(21.6_{\textrm{enr}}+6.2_{\textrm{nat}})$~\kgy\ of germanium
data was recorded. During Phase~I the muon veto was only shut down during
pauses in the germanium data taking and hence there is no additional loss of
exposure due to the veto.

The two data streams were correlated by using the timestamps of the
events. Prior to Phase~I both DAQ systems were operated with their own
internal clock which permitted undesired jumps in the time offset between the
two systems. For Phase~I both DAQ systems were equipped with the same GPS
timing system so that events can be correlated via the timestamp with high
precision. The length of the germanium trace of 160~\mus\ is taken as a
coincidence window. Most interactions between muons and the germanium array
happen within $\pm$10~\mus, however delayed interactions cannot be
excluded. The germanium DAQ is described in detail in Ref.~\cite{gerda_tec}.

Both systems were physically and electronically very stable over time. After
the deployment of the new BEGe detectors in the second half of Phase~I the
set-up of operational modules was unchanged for the rest of the data taking
with four BEGes, six \geenr\ and one \genat\ coaxial detector. In this period
the mean rate was $r_{\upmu}=(4.01\pm0.04)\times 10^{-2}$/s for the veto (no
cuts applied), $r_{\textrm{Ge}}=(2.87\pm0.06)\times 10^{-2}$/s for all
germanium detectors and $r_{\textrm{coin}}=(9.5\pm0.6)\times 10^{-5}$/s the
rate of physical coincidences.  Due to the low rate of veto and germanium
random coincidences are negligible compared to the true coincident rate.
\subsection{Muon rejection efficiency}
A muon rejection efficiency (MRE) can be obtained by defining a cut for
clearly identified muon hits in the germanium detectors and testing for
coincident veto signals. The rejection efficiency $\varepsilon_{\upmu r}$ is
given as the ratio of these events which are vetoed in comparison to the
entire set. The following cuts were applied to the germanium events of the
\gerda\ Phase~I data to identify muons: either a single hit showed an energy
depostion of more than 8.5~Mev or the summed energy of a multi-hit exceeded 4
MeV. This cut excluded energy depositions from the U and Th decay chains.  In
addition, the germanium test-pulse and quality cuts were applied, but no muon
veto cuts. Out of the 848 candidate muon events identified according to energy
release and multiplicity in the Germanium detectors, 841 are actually in
coincidence with a valid muon veto signal. This leads to a MRE of:
\begin{align}
\varepsilon_{\upmu r}=(99.2^{+0.3}_{-0.4})\%
\end{align}
This is a conservative number since the studied events are not standard events
either in germanium multiplicity or in energy range where saturation might
set in.

The derived MRE is slightly lower in comparison to the efficiency derived from
the simulation given in Eq.~\ref{eq:simeff}. Assuming that the given MRE can
be projected to standard events, i.e. with multiplicity $m=1$ and with
energies at or around the ROI, the rejection power of the veto is reliably
high and close to unity.
\subsection{Muonic background index}
\label{subsec:coincsp}
In order to estimate the improvement of the background index given in
cts/(keV$\cdot$kg$\cdot$yr) due to the muon veto, a $\pm100$~keV window was
chosen around \qbb. Analog to the germanium background a blinding window of
$\pm20$~keV around \qbb\ was omitted from the analysis hence the ROI of this
study is 160~keV wide. Out of a total exposure of \exposure=27.8~\kgy\ of
germanium data, 14 vetoed events were found in this ROI with a germanium
multiplicity of one. Were these 14 events not vetoed, they would have led to a
contribution to the BI of:
\begin{equation}
  \textrm{BI}_{\upmu}=(3.16 \pm 0.85)\times \textrm{\pIIbi}~.
\end{equation}
A simulated value for the muonic background in the germanium array surviving
anti-coincidence cuts is~\cite{magemu}
\begin{equation}
\textrm{BI}_{\upmu}(\textrm{MC})=(1.6\pm0.1)\times \textrm{\pIIbi}~. 
\end{equation}
As this simulation was undertaken before construction of the experiment was
finalized the geometry differs to what was realized. Due to the different
geometry, the low statistics and the subsequent large errors, both results can
be considered sufficiently in agreement.

With BI$_\upmu$ and the previously derived MRE, an estimation of the unvetoed
background contribution can be given. It is assumed that the MRE is constant
over the entire energy range of the germanium detectors. The given vetoed BI
is equivalent to the amount of successfully vetoed muons, i.e. 99.2\%. An
amount of unvetoed muons is found:
\begin{equation}
\textrm{BI}_{\upmu,\textrm{unvet.}}=(2.87 \pm 0.77)\times 10^{-5} \textrm{\ctsper}~.
\end{equation}
The design goal of Phase II of \gerda\ aims to reach a total BI of \pIIbi. Thus,
with the current settings of the muon veto unchanged, unvetoed muons would
contribute 1/40 of the BI ``allowance''. For Phase~I of \gerda\ this is
equivalent to 0.16 events in a 200~keV analysis window.

\subsection{Panel detection efficiency}
In order to determine the efficiency of the panels, a data sample of clearly
identified muons was selected. A cut on the Cherenkov events of $M\geq$20 was
chosen which discards unphysical events at low $M$.  The pre-selection of muon
events by the Cherenkov veto is necessary because the standard trigger of the
plastic veto can not be used. For the panels a simple cut on the pulse heights
is insufficient due to the \ga\ coincidences at low pulse heights (see
Fig.~\ref{fig:panscat}). In addition to the pulse height cut defined by the
trigger threshold, a cut on the product of two pulse heights in the form of
Eq.~\ref{eq:eq1} was applied. The detection efficiency of one panel can now be
given as the ratio of events, in which the top and bottom panel of a stack
have fired in comparison to the events where all three panels in a stack have
fired.

The data set contained events since the beginning of the muon data taking with
the panels in August 2011. In this set, 30~044 events were found which
triggered the top and bottom panel. Of these events 29~951 triggered the third
panel as well. This leads to an average muon detection efficiency per panel
of:
\begin{equation}
\varepsilon_{\upmu d}^\mathrm{P}= (99.70^{+0.03}_{-0.05})\%
\end{equation}
Since this value is an average over different panels, it can be seen as an
approximation for a general panel efficiency. The efficiency of a triple stack
of panels is hence $(\varepsilon_{\upmu d}^\mathrm{P})^3 =
(99.10^{+0.09}_{-0.15})$\%. Due to insensitive areas at the panel borders
(scintillator edges, encasing) the effective area of the panels is reduced by
$\sim$5~mm per border or less than $<0.25$\% of the area.
\subsection{Low multiplicity enhancement \label{sec:bump}}
The enhancement at low multiplicities is characterized by a very low number of
recorded photo-electrons (in most cases one or two p.e. per PMT) without any
observable clustering effects and amounts to about 8.7\% of the overall
Cherenkov activity (i.e. $3.1\times10^{-3}$/s). This behavior suggests a very
faint source of light inside the \WT\ that is not directly caused by
muons. Events which trigger the panel veto and can hence be considered true
muon events do not show this anomaly. It was already suggested that this
anomaly is caused by scintillation of the reflective foil under irradiation by
$\alpha$ sources in the stainless steel of the
\WT~\cite{gerda_tec,ritter_phd}. The foil has a highly reflective front side
and is covered with an adhesive on the back side. According to the
manufacturer the foil itself has a thickness of 66~\mum\ and the adhesive a
thickness of 38~\mum~\cite{vm2000}. If the adhesive can be assumed to be an
organic compound a mean free path of $\sim$30~\mum\ for a 5~MeV $\alpha$
particle is expected.  Hence it is unlikely that $\alpha$ particles coming
from the back side are able to deposit energy in the foil and produce
scintillation photons which can exit the foil on the front side. This was
tested by illuminating the foil with an $^{241}$Am $\alpha$ source and by
recording the scintillation light on the front side with a 9235Q PMT. When the
foil was illuminated from the back side, i.e. through the adhesive, almost no
additional light was recorded. When the adhesive was removed from the back
side the same measurement yielded a small effect. If illuminated on the front
side sufficient photons were recorded which could explain the enhancement if
applied to the conditions in the \WT. However, this suggests that the $\alpha$
source is either solved in the ultra-pure water or adhering to the front side
of the foil. The activity of water from this plant was measured to have an
overall $\alpha$ activity in the range of
10$^{-6}$--10$^{-7}$~Bq/kg~\cite{water} and measurements of \gerda\ water
samples agree with these values~\cite{gerdawater}. This activity is too low to
explain these excess events. If the front side of the foil had a radon
contamination, a higher rate would have been expected after the opening of the
\WT. A higher rate was indeed measured but this was in accordance to a higher
dark rate after prolonged exposure to light of the Cherenkov PMTs.
 
\begin{figure}[b!]
\begin{center}
\includegraphics[width=.48\textwidth]{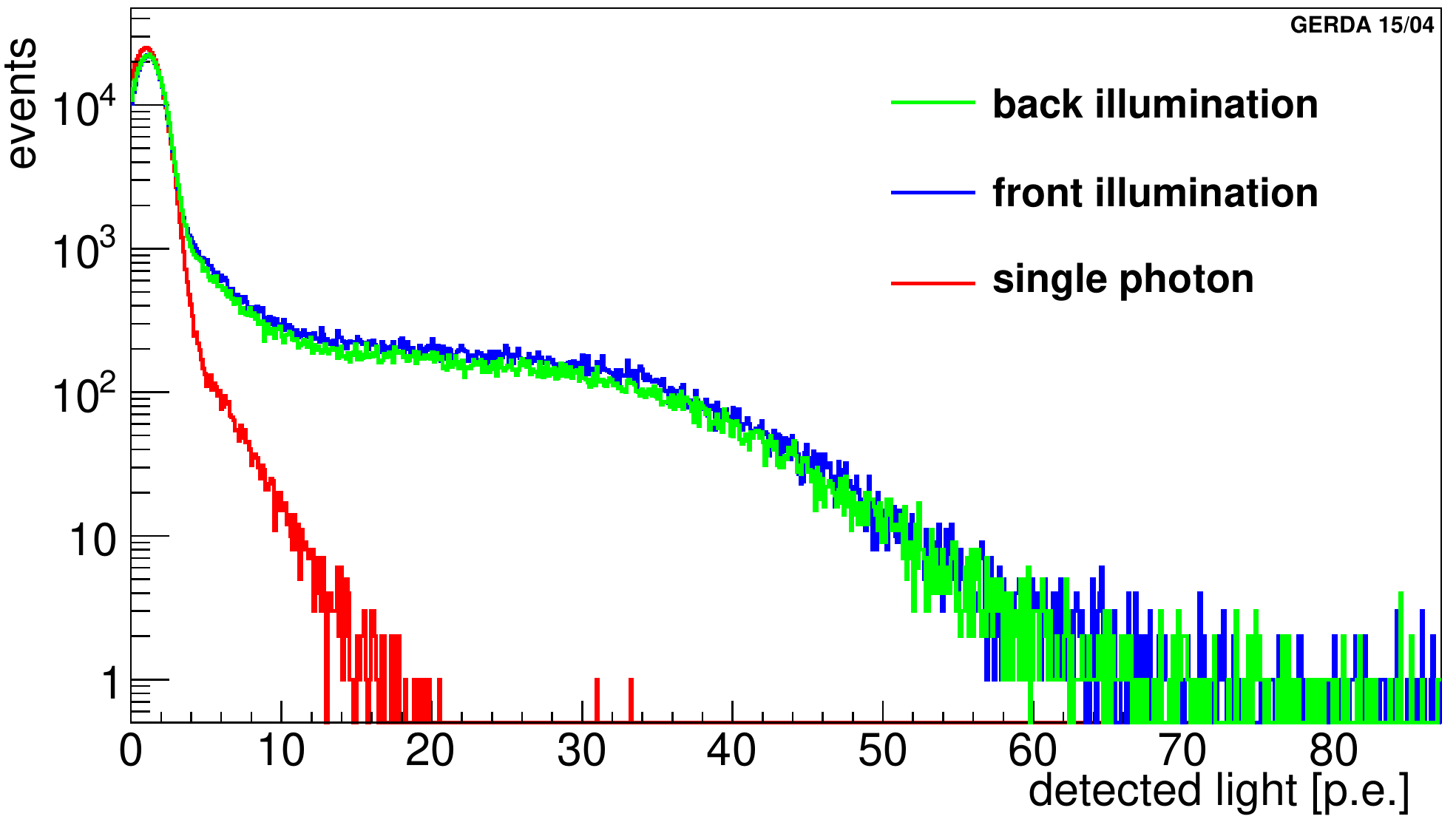}
\caption{    \label{fig:vmfrontback}
     The effect of irradiation of the VM2000 reflective foil by a
     $\beta$ source ($^{60}$Co) recorded by a PMT. The single photon response
     (red) is compared to an irradiation from the front (blue) or back side
     (green) which carries the adhesive
}
\end{center}
\end{figure}

Another explanation for the origin of this anomaly are $\beta$ sources since
electrons have much higher specific ranges in comparison to
$\alpha$ particles. The stainless steel of the \gerda\ \WT\ exhibits 
a low level of radioactivity and one of the strongest contribution to
the radionuclides in the steel is the $\beta$ emitter $^{60}$Co with an
activity of $\sim$20~mBq/kg~\cite{gerdasteel} which leads to a $^{60}$Co
activity on the surface of the \WT\ of $\sim$5~Bq. To test the effect the foil
was illuminated by a $^{60}$Co-source ($Q_{\beta}=0.35$~MeV) and the results are
shown in Fig.~\ref{fig:vmfrontback}.
The foil was illuminated from
the front and from the back side (adhesive not removed). The back
illumination shows only a slightly smaller scintillation effect in comparison to
the front illumination and in both spectra a low-energy $\beta$ spectrum
emerges. In order to determine an efficiency for this process, the scintillation
rate of a 5~mm thick sheet of plastic scintillator was recorded that is assumed
to have an efficiency of unity. By comparing the rate of the foil with the rate
of the scintillator, the efficiency of the foil towards $0.35$~MeV
$\beta$ particles can be calculated:
\begin{equation}
\varepsilon_{\beta}^\mathrm{foil}= (12.0^{+1.1}_{-1.0})\%.
\end{equation}
With this efficiency, the activity expected from $^{60}$Co from the steel is
reduced to $\sim$0.6/s which is still too high in comparison to the measured
rate. Using the mean light recorded by the illumination measurements, the
solid angle (8.5\% of 4$\pi$) and efficiency (0.25) of the PMT used for this
test, the efficiency (0.3) and surface coverage (0.005) of the PMTs in the
\gerda\ \WT\ the effect of this scintillation can be calculated. It is found
that the bulk of these events would deposit 2-6 p.e. in the PMTs per event.
These events will on average not fulfill the trigger conditions given in
Sec.~\ref{sec:daq} It is still likely that about 1\% of these events could still
trigger the veto. This would put the expected and measured rate as well as the
expected photon yield of one to two p.e. per PMT into the same order of
magnitude. Thus, the scintillation caused by $\beta$ particles has to be
considered the most likely source of this enhancement.
\section{Summary}
In this work, the muon veto deployed during Phase~I of the \textsc{Gerda}
experiment was introduced,  its hardware was presented and its performance
was shown. In addition, the cosmogenic components of the background in
\textsc{Gerda} was systematically identified, analyzed and compared to Monte
Carlo simulations.

The hardware thresholds were chosen in a way, that by design a very pure muon
sample is recorded with only a few percent contamination by random
coincidences or other sources of background. With this powerful muon veto over
two years of data have been recorded including the 491~days coincident with
the germanium detectors.

The Monte Carlo simulations of earlier works were extended to accommodate for
a more realistic set-up of the muon veto. PMT multiplicity and p.e.  spectra
were found in good agreement with the data and the light deposition of
cosmogenic and energy-depositing muons could be related to their different
track-lengths in the water tank. With the updated geometry a detection
efficiency of $\varepsilon^{\textrm{sim}}_{\upmu d}=(99.935\pm0.015)$\% was
found for the simulated energy-depositing muons in the \WT.

The muon events in the germanium detectors were studied in detail. A cut on
the signals of the germanium detectors was defined that identifies muon
hits. From comparison to coincident muon events a rejection efficiency of
$\varepsilon_{\upmu r}=(99.2^{+0.3}_{-0.4})$\% was found. The events would have
produced a background index of
$\textrm{BI}_{\upmu}=(3.16\pm0.85)\times\textrm{\pIIbi}$.
\section*{Acknowledgments}
The work of the T\"ubingen group within the \gerda\ experiment is supported by
the grants 05CDVT1/8, 05A08VT1, 05A11VT3 and 05A14VT2 provided by
the BMBF/PT-DESY.

Special thanks goes to the mechanical and electronic workshops in T\"ubingen
for their assistance during the preparation of the Cherenkov veto. M. Uffinger
assisted in the activity measurements of the VM2000. The authors would like to
thanks B. Schwingenheuer for useful discussions and for carefully reading the
manuscript.


%
%
\end{document}